# NSF Integrated Circuit Research, Education and Workforce Development Workshop Final Report[1]

## Steering Committee

- Chair: Matthew Guthaus <mrg@ucsc.edu>
- Christopher Batten <cbatten@cornell.edu>
- Erik Brunvand <ebrunvan@nsf.gov>
- Pierre-Emmanuel Gaillardon <pierre-emmanuel.gaillardon@utah.edu>
- David Harris <harris@g.hmc.edu>
- Rajit Manohar <rajit.manohar@yale.edu>
- Pinaki Mazumder <pmazumde@nsf.gov>
- Larry Pileggi <pileggi@cmu.edu>
- James Stine <james.stine@okstate.edu>

## Executive Summary

As the pace of progress that has followed Moore's law continues to diminish, it is critical that the US support Integrated Circuit (IC or "chip") education and research to maintain technological innovation. Furthermore, US economic independence, security, and future international standing rely on having on-shore IC design capabilities. New devices with disparate technologies, improved design software toolchains and methodologies, and technologies to integrate heterogeneous systems will be needed to advance IC design capabilities. This will require rethinking both how we teach design to address the new complexity and how we inspire student interest in a hardware systems career path. The main recommendation of this workshop is that accessibility is the key issue. To this end, a National Chip Design Center (NCDC) should be established to further research and education by partnering academics and industry to train our future workforce. This should not be limited to R1 universities, but should also include R2, community college, minority serving institutions (MSI), and K-12 institutions to have the broadest effect. The NCDC should support the access, development, and maintenance of open design tools, tool flows, design kits, design components, and educational materials. Open-source options should be emphasized wherever possible to maximize accessibility.  The NCDC should also provide access and support for chip fabrication, packaging and testing for both research and educational purposes.

---

[1] This material is based upon work supported by the NSF under Grant No. 2137629.

# Table of Contents

# Introduction

Chip design at US universities blossomed on account of the Mead and Conway Revolution (Conway). In 1978-79, Carver Mead and Lynn Conway wrote the seminal textbook *Introduction to VLSI Systems* [1]. This book offered abstractions that transformed digital chip design from a complex physics problem into a much easier computer science problem and popularized chip design in academia.

Conway also taught a VLSI course at MIT in 1978 leading to the Multi-Project Chip concept, and Danny Cohen established the Metal Oxide Semiconductor Implementation Service (MOSIS) at USC for VLSI prototyping. MOSIS fabricated free chips for university VLSI classes, initially with NSF support and later with profits from their commercial operations, but ceased offering this service in 2020. DARPA also kicked off a VLSI research project in 1980, popularizing Mead & Conway's work and encouraging the development of chips and electronic design automation (EDA) tools.

Chip design was further fueled by a collaboration of universities and industry. Much early design was done with the MOSIS layout tool, but more advanced design requires synthesis and placement & routing tools. EDA vendors, including Cadence, Synopsys, and Mentor Graphics, established university programs with discounted tools. North Carolina State University developed the NCSU Process Design Kit (PDK) with modern predictive (i.e., non-fabricable) technology files [2].

Through the 1990s, universities taught large VLSI classes at the graduate or advanced undergraduate level using a vibrant set of textbooks. Fabricating a chip through MOSIS and testing it became a rite of passage for thousands of engineering majors. Brunvand provided a cookbook to install and configure EDA tools and models for chip design with industry-standard tools on readily available processes [3]. Flows became mature enough to be accessible to all university students.

In recent years, VLSI education has faced a decline leading to VLSI classes now being offered only as a niche topic. According to faculty at our recent NSF-organized panel discussion, enrollments have dropped at most universities and the fraction of underrepresented students in these courses is close to zero at some schools. Textbook sales have sharply declined and predominantly moved to India and China. MOSIS has dropped support for the old 0.6-micron process long used for class projects and dropped class funding entirely. The cost of discounted VLSI EDA tools is significantly higher than other university software and the cost per student increases when enrollments decline. Many universities are proposing to stop purchasing these academic licenses entirely. Besides, the cost of maintaining these tools and their computing infrastructure requires IT staffing and expertise as well as dedicated computer systems for licensing and installation. Licenses prevent students from using tools outside of traditional computing labs, notably on personal computers which are the de facto standard for most academic programs.

VLSI research in academia has become very difficult because results using antiquated processes are generally not publishable and the design files (device models, design rules, and libraries) for advanced processes are proprietary and available only to faculty with special industry connections. Often universities are unwilling to sign the Non-Disclosure Agreements (NDAs) that some industrial partners require due to numerous restrictions on IP ownership, export control, liability, lock outs, and other challenges. The cost of fabrication in advanced processes is also so high that it often requires leveraging an existing shuttle run from a corporate sponsor, again only available to faculty with special connections. Altogether, far fewer students are graduating from US universities who are prepared for careers in chip design.

The US is now facing both an economic and national security threat in the semiconductor industry. There is presently a worldwide semiconductor shortage that forced Apple to delay the iPhone 12 by two months ("Apple unveils new 5G iPhone 12 line in multiple sizes") and cost the automotive industry a forecasted $60B in revenue [4]. In 2014, Broadcom found itself unable to compete in the cell phone application processor market against foreign competition from Samsung and MediaTek, closed the entire division, and was weakened to the point that it was acquired by a Singapore company, Avago [5].

Because of limited supplies, there is a multibillion-dollar black market in counterfeit electronics and an estimated 15% of spare electronic parts purchased by the US Defense Department are counterfeit, threatening both reliability and security [6]. In 2020, the Federal Communications Commission designated Chinese telecommunication firms Huawei and ZTE as national security threats because of the risk of espionage through their 5G networking equipment [7]. The Chinese National Integrated Circuit Plan seeks to develop leading-edge domestic integrated circuit manufacturing capability by 2030, and the country invested $150B in its industry between 2014 and 2020 [8]. In 2014, IBM sold their chip business to GlobalFoundries, ending a source of domestic chip manufacturing.

Several forces are coming together with prospects to revitalize the US semiconductor industry. In March 2021, Intel CEO Pat Gelsinger announced a $20B investment to build two new IC fabrication plants in Arizona for 7nm manufacturing and to sell foundry services to outside customers [9]. This will significantly expand US manufacturing capacity. President Biden's "American Jobs Plan" of March 2020 calls for over $50B investment in domestic chip manufacturing [10]. The critical challenge, however, is how we will educate and establish a workforce to create designs for manufacturing in these fabs.

There has been a significant interest in how open-source hardware designs and EDA tools can impact IC design, which has been notoriously closed-source. DARPA has funded broadly the development of an entirely open-source toolchain [11], and open-source Field Programmable Gate Array (FPGA) tool flows have also seen very mature toolchains for multiple FPGA architectures [12]. DARPA has also funded the Electronic Resurgence Initiative to secure the supply chain and promote integrated circuit research and development [13]. The Free and Open Source Silicon Foundation (FOSSi) is promoting open-source EDA tools and libraries [14]. Industry consortia such as CHIPS Alliance have coalesced around such open-source opportunities [15]. Google has recently begun sponsoring SkyWater Multi-Project Wafer shuttles for any open-source project in the older SkyWater 130nm technology and recently announced plans to develop an open source process design kit (PDK) for SKY90-FD, SkyWater's commercial 90nm CMOS process technology [16] as well as a GlobalFoundries 180nm technology [17]. One monumental success in the open-source area has been the adoption of RISC-V by numerous companies and academic institutions [18]. Entire microprocessor ecosystems, which were once filled with proprietary ISAs, firmware, and compilers, are now freely available and competitive in performance.

While the Mead-Conway revolution began by moving IC design from analog designers to digital designers, the open-source movement may enable a shift from digital designers to software engineers for the next generation of "IC designers." Hardware design has benefitted from many software design efficiency and productivity paradigms such as high-level synthesis, intermediate representations, programming languages, formal verification, etc. Recently, for example, new high-level design languages such as Chisel [19] and XLS [20] have started to change how we abstract hardware design and offer new opportunities for optimization and integration [19]. Improved accessibility of open-source IC design ecosystems opens the opportunity for innovation as well as a chance to increase diversity by stretching the concept of who can be an IC designer.

As Moore's law wanes, it is even more critical that universities reconsider their approach to IC education and research. Improvements in IC designs will need to come from new devices with disparate technologies, improvements in EDA toolchains and methodologies, and integration of heterogeneous systems. All of these will require rethinking both how we design to address the complexity and how we revitalize student interest in hardware systems. Academics and industry must address this change now when training our future workforce.

In the last two years, there have been related workshops on similar problems [21], [22]. In particular, these focused primarily on manufacturing and research. While this was also a part of our workshop, we took a broader consideration of workforce development which includes education and research at all levels: K-12, undergraduate, and graduate (MS and PhD).

As a part of this workshop, we held two meetings to discuss the issues. An initial virtual workshop was held on October 14-15, 2021 with a public invitation. This virtual workshop had over 70 participants from academia, industry, and government that were very active for the entire two days (see Appendix A). This workshop resulted in a focus of topics (see Appendix B) which guided the selection of a short-list of participants for an in-person workshop. The in-person workshop was held on May 20-21, 2022 at the UC Santa Cruz Silicon Valley Center and had 33 registered participants (see Appendix C). While a few were unable to attend due to COVID-related issues, those present spent the two days in panel discussions, breakout sessions and then beginning an initial draft of this report (see Appendix D).

# Technology Nodes

The chip design community relies on a broad spectrum of manufacturing technology nodes (or "feature sizes"). This is reflected in the charts below, which break down prototyping fabrication service bookings for the Europractice fabrication service in 2021 and MOSIS in 2018-2022. The choice of technology depends on a variety of factors, which are examined in more detail below. In addition to the technology node, there are a number of other features leading to an even wider diversity. For example, besides basic logic CMOS, there are technology flavors including add-ons for radio frequency (RF), high-voltage (HV) circuits and non-volatile memory.

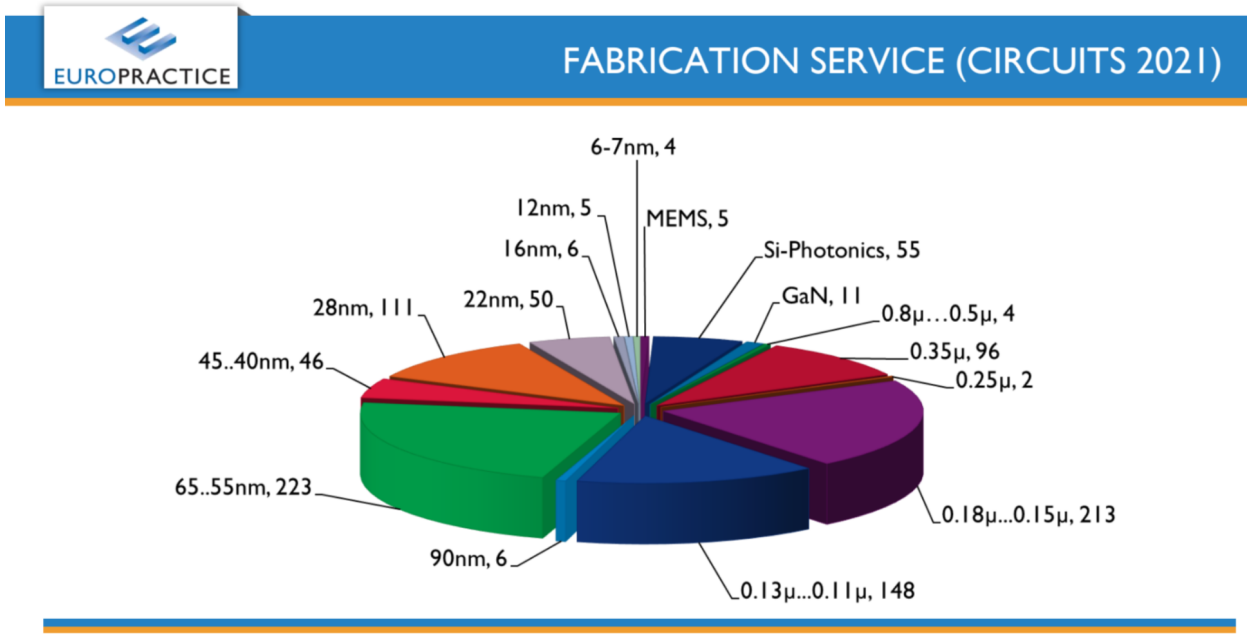

Source: Romano Hoofman, Europractice, 2021.

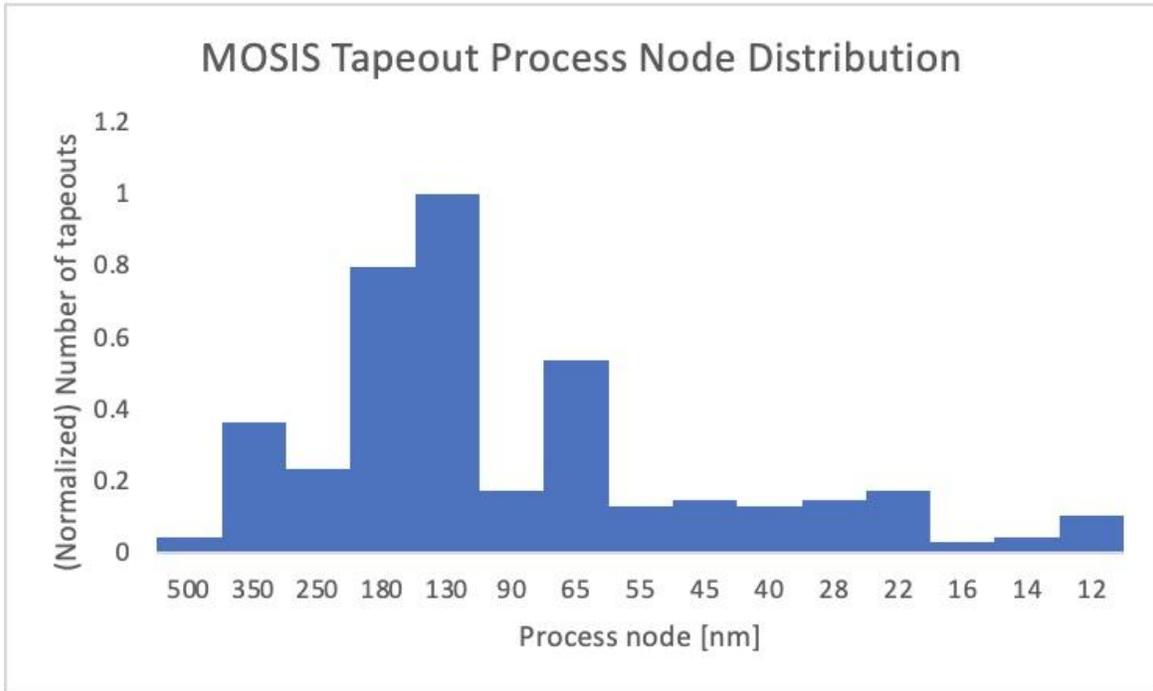

Source: Lifu Chang, The MOSIS Service, January 2018-May 2022.

Relatively old technology nodes (40nm and above), make up nearly three quarters of the bookings. This is partly explained by not all applications requiring small feature sizes, such as low-speed sensor applications. The usage of these trailing-node technologies is not limited to academic exploration, but is also high volume in a variety of commercial applications that are analog-centric. For education, older nodes are attractive since they provide an easier entrance to the field and do not have the cost and other complexities associated with newer nodes.

Recently, Google and SkyWater Technology have open-sourced SkyWater's 130nm technology, which has become an attractive option for researchers, educators, and even hobbyists. This technology is good enough for designing a wide variety of useful analog, mixed-signal and RF circuits as well as digital processors and accelerators with moderate performance. The 130nm node even contains options for non-volatile SONOS (silicon-oxide-nitride-oxide-silicon) flash and $HfO_x$-based Resistive RAM (ReRAM) devices. Even more recently, Google and SkyWater have announced the same plans for its 90nm technology [23] which will bring additional performance opportunities. There is also a plan for another trailing-node 180nm technology [17] from GlobalFoundries.

Newer technologies are needed to push the boundaries in high-performance digital and wideband analog/mixed-signal/RF circuits. The state of the art (SOTA) is usually defined as the FinFET/SOI processes and onward which are offered by foundries on 22nm, 16nm, 12nm and smaller-feature-size processes. We note that the ecosystem around advanced nodes (7nm and below) support the R&D of Artificial Intelligence (AI) and Machine Learning (ML) IC designs. Therefore, research projects in these areas usually receive full support of not only the fabrication, but also the ecosystem of IP, EDA, software, etc. Because of this, there has been an increasing demand from universities for access to SOTA processes. In particular, this has been seen in DARPA and AFRL joint sponsored projects as well as with individual university research projects. In the last five years, there have been about thirty university design projects at 22nm and 16nm. However, not only are there more university teams intending to adopt the SOTA processes, but they also should be provided with access to the more advanced nodes at 7nm, 5nm, and onward.

The capabilities of logic CMOS process technologies are enhanced by adding additional active and passive devices, as well as optimizing the interconnect technologies. Processes with such enhancements are widely used for research. For instance, SiGe HBTs are added to improve power handling and high speed and frequency performance; high voltage devices are added to support the necessary voltage for power management applications; and phase-changing material (PCM) devices are added to enable post fabrication programming. A wide variety of memories such as DRAM, Flash, ReRAM, STTRAM are also included to satisfy varying storage needs. The memories are also being used for in-memory as well as mixed signal computing research. The PCM and MEMS devices are also added to support low-loss switch applications. MEMS devices are also integrated with CMOS systems to support a wide range of sensing applications. Another important research area is silicon photonics, which utilizes integrated circuits that incorporate both CMOS and photonic components for sensing and communication.

Open PDKs use representative, predictive or actual technology parameters to enable research and teaching without many of the challenges associated with proprietary PDKs. Some of these are not fabricable such as NCSU's FreePDK45 [2], NCSU's FreePDK3 [24, p. 3], ASU's ASAP7 [25], Synopsys's SAED 90/32, and Cadence's GPDK 180/45/7. More recently, we have begun to see fabricable, open-source PDKs such as the Google/SkyWater 130nm PDK [26] and the forthcoming SkyWater 90nm and GlobalFoundries 180nm PDKs [27]. Open PDKs are either freely available with permissive open-source licenses or at least easily accessible without cost to academic researchers. These open PDKs can be useful for setting up EDA tool flows, exploring EDA algorithms, architecture design-space exploration, and/or teaching IC design courses, but do have some trade-offs. The PDKs from Synopsys and Cadence only work with proprietary tool flows whereas the NCSU options have seen broad adoption in both open-source and proprietary EDA flows. While ASAP7 is gaining support in both open-source and proprietary EDA tool flows, it was originally restricted to non-commercial usage which limited its early adoption in industry-sponsored research.

The needs for specific technology nodes and access services can vary significantly based on whether one is pursuing a tape-out for research or for teaching. Research needs can span the entire spectrum of technology nodes from the oldest (180nm or larger) to the most recent (5nm or smaller) depending on the kind of research (e.g., digital/analog/mixed-signal circuits, architecture and system-on-chip design, emerging devices), the target research metrics (e.g., performance, energy, form factor, fault tolerance), and ultimately the research question that is being addressed. Teaching needs are fundamentally different and focus more on the educational experience as opposed to pushing the state-of-the-art. Teaching can often use older technology nodes to reduce cost, complexity, and legal challenges, but this currently fails to provide many real-world experiences required by more recent technology nodes (e.g., complex design rules and signoff requirements, larger parasitics, increased variability and noise, complexity of relevant designs, etc.).

## Summary of Current State

Currently, there are two approaches for instructional tape-outs: industry subsidized in relatively recent technology nodes (e.g., TSMC 28nm at UC Berkeley and CMU) and low-cost tape-outs on older technology nodes (e.g., SkyWater 130nm through Efabless at Cornell, Yale, Stanford, and UCSC). There is also the third option where a design is not taped-out at all. The following are providers of fabrication services, usually for a fee:

- **MOSIS:** The MOSIS Service ([www.themosisservice.com](www.themosisservice.com)) is a for-fee fabrication service provider for universities through GlobalFoundries, TSMC, Intel Foundry Services, Samsung Foundry, and WIN Semiconductors foundries. The portfolio of the silicon process technologies ranges from 350nm to 12nm FinFET, with specialty BCD (Bipolar-CMOS-DMOS) and other processes as well as Magnetoresistive RAM (MRAM). The portfolio also includes III-V compound semiconductor GaAs and GaN processes. The projects can be either MPW or dedicated (full mask) runs. MOSIS provides PDK files, training, design debug, design milestone checks, and tape-out sign-off and other necessary support items to its paying customers. MOSIS previously had a free academic program for tape-outs which was supported by NSF and industry but was discontinued in 2020.
- **MUSE:** MUSE is a for-fee, multi-project wafer service founded in 2018 with a current focus on TSMC shuttles ([https://www.musesemi.com](https://www.musesemi.com)). MUSE originally focused exclusively on university researchers but has recently broadened its scope to supporting commercial companies. MUSE provides shared block access to TSMC 180nm, 65nm, 40nm, and 28nm with general logic, low power, and RF variants. MUSE has small minimum die sizes (5mm$^2$ in TSMC 180nm MS-RF-G, 5mm$^2$ in TSMC 180nm in HV-BCD-G2, and 1mm$^2$ in smaller technology nodes). MUSE also supports full block and production tape-outs on all TSMC technologies. MUSE can also provide access to the corresponding PDKs, standard cells, I/O cells, and SRAM compilers.
- **EUROPRACTICE:** EUROPRACTICE ([https://europractice-ic.com/](https://europractice-ic.com/)) has supported academia and industry with for-fee MPW prototyping services, training activities, system integration solutions and small volume production for over 25 years. EUROPRACTICE provides academic institutions and start-ups in Europe with access to design tools. EUROPRACTICE provides a full suite of technologies including CMOS technologies with feature size ranging from 2um to 12nm, SiGe HBT BiCMOS technologies, and for photonics, MEMS, microfluidic, graphene, SiC and others.
- **CMC:** CMC Microsystems ([https://www.cmc.ca/fab/](https://www.cmc.ca/fab/)) provides for-fee MPW services primarily for Canadian institutions. It supports a variety of CMOS and BiCMOS technologies, Silicon photonics and MEMS processes, as well as III-V Epitaxy.

- **Efabless:** Efabless ([https://efabless.com/](https://efabless.com/)) provides an ecosystem for community chip creation which includes prototyping through low-volume production using a MPW service for mature to mid-range technology nodes. Recently, Efabless has become the de-facto standard for accessing the open-source SkyWater 130nm PDK through open-source and proprietary EDA tool flows. Efabless supports the Google-funded OpenMPW program which provides opportunities for students, researchers, professionals, and even hobbyists to tape-out a chip on SkyWater 130nm at no cost as long as the entire design is fully open source and is reproducible with open-source EDA tools. Efabless also provides the for-fee ChipIgnite program for a guaranteed MPW slot and no open-source requirement. Efabless provides turn-key container images with the open-source SkyWater 130nm PDK, open-source EDA tool flow, and Caravel SoC harness to simplify the design and tape-out process. The solution includes pre-designed packaging, a standard PCB evaluation board, and test software to streamline chip bring-up and test.

## Key Challenges

**Fabrication Costs are Significant**

For the latest technology nodes, the key challenge is cost, not only pertaining to fabrication, but also managing design and verification complexity, as well as usage and maintenance of designs kits and tools. Even for some older technology nodes, fabricating designs with a reasonable number of transistors can quickly exceed many tens of thousands of dollars. For very old technology nodes, each individual tape-out cost might be less, but having to fabricate many different unique designs for a course can quickly add up. It is important to recognize that universities have drastically different internal resources to support fabrication costs and the limited budget per student for one class is much less than fabrication costs. Even universities with well-established programs in IC design have to leverage industry connections to fund instructional tape-outs.

Yet another challenge related to cost comes from minimum die size requirements. In FinFET technologies (typically 16nm and below), some services require a minimum chip area of 2mm X 2mm, which comes at a cost close to $100,000.

**Limited Shuttle Space**

The shuttle programs can be overbooked and sometimes can be canceled. In the case of the OpenMPW shuttles provided by Google and supported by Efabless, a lottery system is used to select which set of 40 projects is fabricated on each shuttle. Interest in particular technology nodes, such as older ones, may be mostly desired by academia yet not desirable enough to make economic sense for a fabrication run. These issues can have a significant impact on research progress and student graduation.

**Legal Agreements to Access Advanced Technology Nodes are Complex**

Access to advanced technology nodes is often governed by complex NDAs and other usage restrictions. For example, some foundries require a one-year "cooldown period" for a user who wants to switch from their technology to a comparable offering by a competitor. In addition, export control must be considered for the latest nodes, often leading to complex procedures and non-inclusive access on university campuses for different students. In some cases, this has even required dual-flows in a single class for using commercial and a non-fabricable open PDK depending on student national origin and visa status, or even current location (within US borders versus outside while traveling).

**Limited Access to PDKs**

Because many foundries require a commitment to tape-out before a PDK is provided, each design team must have a determined/funded tape-out project before they can acquire a design kit. In cases where a commitment is not required, PDKs with limited information may be provided until the tape-out commitment is made. This requirement is legitimate because the foundries would like to have a clear PDK procedure that will guarantee a project and will not be a waste of time or risk of IP leaks. However, this brings an immediate chicken-and-egg challenge - without a PDK to evaluate, how can a team determine a tape-out project? The immediate next challenge is that the team may need to do test-chip tape-outs to different processes to determine the selection of the process, which is usually not feasible with their budget and resources.

**Limitations of Non-Manufacturable Open PDKS**

While non-fabricable open PDKs can be useful for setting up EDA tool flows, exploring EDA algorithms, design-space exploration, and teaching IC design courses without a tape-out component, they are not sufficient for realistic analog or mixed-signal research. The accuracy of these open PDKs can vary significantly and since they are not based on a real technology node, there is not a clear path to tape-out. Newer, fabricable, open PDKs such as the Google/SkyWater 130nm address this problem, but do not provide access to the most recent technology nodes.

**Installing and Maintaining PDKs is Challenging**

PDKs are complex and very large in size (1TB+) especially for the advanced nodes (28nm and lower). Installation of PDKs can take multiple days by specialists. Often problems are encountered and resolving them takes multiple email exchanges with the foundries or MPW aggregators spanning many days. These PDKs also require updating multiple times per year.

**High-School Programs and Community Colleges are Often Ignored**

High-school students are the IC designers of tomorrow, yet high-school students currently have very little (if any) access to learn and experiment with IC design. Community colleges can provide critical vocational training for the IC design industry, yet the challenges associated with IC design mean that very few community colleges have any kind of IC design course work or tape-out opportunities.

## Recommendations

**NSF Support for a National Chip Design Center to Manage Research and Instructional Tape-outs**

There is a critical need for a National Chip Design Center (NCDC) to serve as a centralized entity that provides the "three pillars" of design enablement to universities: training, EDA tools, and fabrication service:

(1) Training: The NCDC should provide in-depth and detailed training for university design teams about the best practice towards a successful tape-out targeting each foundry process. The scope of a training should include the PDK, IP, EDA tool, chip integration, packaging bump design, foundry cell insertion, design milestones and checklist, tape-out sign off requirements and checklist, and fabrication schedules. The training should also include reliability, ESD, yield optimization, and other nanometer design issues. The training may need to be offered multiple times per year or accessible online.

(2) EDA tools: EDA tool support is critical and most university programs do not provide support or have limited support options. The NCDC would provide known tool flows complete with licenses and version requirements as well as support to minimize the EDA tool bugs and errors. One way is to support a reference run script that utilizes a reference design. This reference kit demonstrates the essential commands for successful and accurate runs. Support (of the reference script and the reference design) will help the designers learn the EDA process and reduce the need of supporting specific EDA errors due to setup, versioning, and misuse. Because the designers know the basic runset versus their own, additional commands, they can do more self-guided debugging. For the highest accessibility, the reference tool flow and design should be open source so that they may be shared, modified, scaled to explore the design implementation space, and not be restricted from redistribution.

(3) Fabrication Service: The NCDC should provide fabrication services for both research and instructional purposes. This would include MPW aggregation, dicing, and packaging.

While MOSIS, EUROPRACTICE, Muse, CMC, and Efabless all provide similar services, they are ultimately for a fee or limited in capacity in the case of the Google/Efabless OpenMPW. Therefore, these focus primarily on industry or academic projects with funding.

The NCDC should restart and manage a free instructional tape-out program in which students in community colleges, 4-year universities, and graduate students can tape-out their class design projects. NSF should also enable tape-out opportunities for high school students participating in integrated circuit design programs. Cloud-based design environments should be available as an option to alleviate the infrastructure challenges. An older CMOS node (e.g., 180nm) should be adequate for most instructional tape-outs. A design quality control process should be included.

**NSF Program for Research Fabrication Cost**

In addition to supporting the NCDC, we recommend a separate funding mechanism for research tape-outs. The NSF Major Research Instrumentation Program (MRI) mostly funds equipment and infrastructure and is not well suited towards tape-outs. NSF also funds shared facilities (e.g., supercomputers and telescopes) and then these facilities provide shared access through a request process. We recommend NSF create a dedicated program for funding fabrication costs associated with NSF-funded research as well as academic instruction. This program would be distinct from research core programs. Advanced nodes, which are at a higher cost, should require merit-based review, whereas older nodes for instruction, which are significantly cheaper, should have a straightforward application pathway.

The CMOS+X pilot effort is an excellent start, but we envision a dedicated program that encompasses traditional CMOS fabrication in addition to CMOS+X [28]. This program should fund fabrication costs associated with a wide spectrum of research including digital VLSI, analog/mixed-signal circuits, computer architecture, system-on-chip design, emerging devices including systems using them, etc. If the program is exclusively through CISE, it should have a focus on fabricating systems, while a program in partnership between CISE and other directorates (e.g., MPS, ENG) could expand the scope to circuits and devices. The program should include multiple explicit tiers (e.g., small, medium, large) as well as research and instructional objectives to (i) ensure that funding a few large fabrication projects on advanced technology nodes does not completely overwhelm the funding of smaller fabrication projects and (ii) ensure that large fabrication projects are not seen as too expensive every funding year. The fabricated systems would become critical research artifacts highlighting the work CISE is doing to advance the state-of-the-art in IC design.

**Infrastructure Development Support for Open PDKs**

Open PDKs have become an essential ingredient for teaching and academic exploration. However, as discussed above, their utility is mainly limited to digital design. The NSF should provide infrastructure development funding to researchers and educators who are interested in extending these open PDKs for usage by the broader community, including analog, mixed-signal and RF designers. This funding could also support packaging, documentation, and reference flows for using these open PDKs. As discussed later in the IP section as well, the development of essential design collaterals (memory compilers, IOs, level shifters, power and clock gating cells, etc.) should also be funded to enable complete design experiences and training based on open PDKs. Development of key IPs (external memory interfaces, on-chip NoCs, etc.) will enhance value and adoption as well. In addition to development funding, the NCDC should be funded to maintain and enhance such infrastructure and associated tooling, train on their usage, and integrate them into the recommended tape-out EDA flows.

**Cloud-based Design Environment**

We also recommend a secure, cloud-based design environment where users can try a set of different PDKs and EDA tools as well as leverage cloud computing resources, which can be significant. The NCDC or another management organization should maintain and run the environment. In the cloud-based trial environment, users can run and then decide which PDK to use towards a tape-out or decide to not tape-out. The user can also perform EDA tool research that must utilize advanced node PDKs. The environment should protect the IP of both the foundries and the design teams. With the environment that is PDK-secure and encouraging trials of different PDKs, eventually there will be more tape-out projects with the foundries which should entice their participation.

# Training and Education

Universities will face substantial challenges in meeting the emerging need for trained semiconductor engineers. By 2030, estimates are that only there will only be enough qualified candidates to fill thirty percent of semiconductor manufacturing jobs [29], [30]. Computer hardware engineers tend to have slightly higher salaries than software engineers, but the growth is significantly less and there are many fewer jobs overall [31], [32]. However, if we consider Electrical and Electronics Engineers, who have training that is required for many semiconductor jobs, the average salary is significantly less than both computer hardware and software engineers [33]. Because of this, semiconductor companies struggle to recruit qualified American candidates compared to their computer and software counterparts, which often leads to outsourcing or hiring of H1B visa workers.

There are several factors contributing to this phenomenon. First, the overwhelming focus of semiconductor education is on MS and PhD programs, which means outreach and recruiting for these programs typically starts in the student's 3rd year in an Electrical or Computer Engineering (ECE) program. As a result, STEM outreach programs for hardware and semiconductor education are virtually non-existent compared to the robust programs for Computer Science and Software Engineering. In addition, a typical American child is given a consumer electronics or mobile device between the ages of 8-11 years old and they gain substantial expertise and understanding of the benefits of the higher levels of abstraction, such as app and website development. The success of integration has led to decreased modularity of modern, portable computing systems which means users have less exposure to hardware systems and associated abstractions.

Unfortunately, many students also often fail to see the benefit of understanding the lower levels, such as Register Transfer Level (RTL) design or Electronic Design Automation (EDA). The successes in IC research and design have abstracted their complexity away from the user, which has made conveying those complexities even harder over a truncated curriculum. A commonly cited concern from industrial leaders is that newly-hired engineers require increasing "on-boarding" times in order to gain familiarity with tools and flows at the advanced nodes. Experiential learning opportunities for students in IC design are extremely limited, in contrast with the rapid growth and deployment of 3D printing in these educational spaces. In order to meet the growing demand, we must develop innovative approaches to engender an appreciation and joy for computer hardware and engineering *substantially* earlier in a student's education.

The barriers for gaining employment in software-related fields are significantly lower than in hardware and semiconductor-related roles. For example, positions in application or web development can be competently filled by candidates even without a bachelor's degree, much less a computer science MS or PhD degree. An equivalent role for a semiconductor technician is virtually non-existent due to the industry's dependence on qualified PhD candidates. As a result, even if the need for semiconductor professionals included the development of a robust technician track, community colleges would be hard-pressed to meet the need without substantial investment, as many community colleges do not necessarily require a PhD to instruct STEM courses. Meeting the critical needs of the semiconductor workforce requires the development of trade programs, as well as recertification programs as integrated circuit developments advance.

There is a crucial need for non-ECE majors to understand the contemporary issues in semiconductor design and manufacturing. Nearly every facet of modern American life is impacted by semiconductors, yet few people outside of ECE-related fields truly understand the gravity of a "Chip Shortage." These issues no longer impact just manufacturing and EDA companies. Experts estimate that the global chip shortage cost the American economy $240 billion in 2021 alone. In addition to science and engineering, students in the fields of business, architecture, law, humanities, arts, social sciences, and global affairs will need to become increasingly literate on contemporary issues in semiconductors. These issues impact all corners of the American economy, which means we require a transformative change to a liberal arts education as well as STEM education.

These semiconductor educational challenges will impact the already-strained capability of the Department of Defense (DoD) to produce mission-critical microelectronics critical to national security. The current state of DoD microelectronics is one of reliance on production of low volume State-of-the-Practice (SOTP) and legacy microelectronics that are unavailable from commercial foundries. DoD identified the challenges of an atmosphere of diminishing domestic semiconductor manufacturing capability and increasing worldwide supply chain risks, which proved prescient prior to the COVID-19 global pandemic. DoD requires a substantial increase in qualified semiconductor engineers in order to mitigate counterfeiting, Trojan horses, specific reliability issues in military environments, and rapid obsolescence coming from an unpredictable and unsecured supply chain. The current US educational system is not prepared to meet these national security challenges.

Finally, the looming "enrollment cliff" adds a specific urgency in addressing the challenges in integrated circuit research and education. By the year 2025, there is an estimated sharp 15% drop in college-aged individuals in the general population due to the 2008 recession and decreased birth rates [34]. This will impact university budgets and strain the already-stressed ECE and CS departments, particularly regional four-year colleges and universities and community colleges. Meeting the emerging needs of the semiconductor workforce presents a critical challenge: it demands that we figure out how to rapidly triple qualified candidates even as there are 15% fewer students and most competing industries have a substantial head start in engaging and inspiring K-12 students.

The volatile mix of limited training opportunities prior to the PhD, resource-depleted universities, a sharp and drastic reduction in eligible college-aged students over the next five years, strained global supply lines, reliance on foreign foundries to meet our semiconductor demand, limited pathways to join the IC workforce, and a population with limited understanding of the contemporary issues is potentially *devastating* for our national infrastructure and defense. Even if the current IC engineering pathway tripled the number of qualified candidates, the educational system cannot meet the emerging demand. This means that diversity, equity, and inclusion issues in integrated circuit research and education *are national security issues*. We require a transformative "all hands on deck" approach to solving these challenges in order to address emerging issues in 2030 and beyond.

## Summary of Current State

The current state of the art in VLSI and semiconductor education is rather bleak. Primary and secondary students, K-12, are not exposed to integrated circuit design concepts and are, at best, exposed to computing as a black box. They do not understand the impact of semiconductors in society. VLSI and semiconductors are under-appreciated; the false perception is that all the innovation is in the software and that the hardware is a solved problem.

In post-secondary education, there are drastically uneven opportunities for VLSI and semiconductor education. There are very few VLSI classes and fewer tape-out (i.e., where students actually make a chip) classes which are only at R1 schools. The majority of students are not in R1 institutions and other institutions (R2 and Minority Serving Institutions) typically do not have programs at all. The connections to other fields, such as robotics, medicine, and computer science, are lacking and the benefits of VLSI and semiconductors to these fields are not often emphasized in post-secondary programs. There are very few students in "hardware" programs because there is a stigma around the field that it is too difficult and has a high barrier to entry. This has resulted in the reduction of classes even at R1 schools. In the past, many R1 schools have had classes in design, testing, and verification, but this has mostly been replaced with a single-semester, elective VLSI course and, as previously discussed, no tape-out opportunities.

The focus of VLSI and semiconductors training and education has been on graduate students in MS and PhD programs, primarily at large R1 schools, but other forms of VLSI engineers, including technicians and application engineers who may not need an MS/PhD, have not been given much attention.

## Key Challenges

**Competing interests in K-12 and early UG**

Increasing student participation in VLSI requires engaging with them at an early age. However, there is a finite amount of time and attention devoted to extra-curricular activities and "specials" education in primary and secondary school settings. Several initiatives have been created at the K-12 level to inspire and motivate students to engage in robotics (e.g., FIRST), coding (e.g, camp K12), artificial intelligence, and physical computing (e.g., Raspberry Pi, Arduino). A key challenge in engaging K-12 students in VLSI is the lack of mature initiatives and an understanding of what is needed to enter this space. There are no studies to understand how gaps in VLSI knowledge of K-12 students can be filled. Additionally, it is unclear whether stand-alone initiatives are needed and/or whether collaborations with existing initiatives will allow for sufficient coverage of VLSI topics.

**Lack of appropriate material and existing material is decentralized**

There are cracks in the information pipeline for K-12, undergraduate, and graduate students who want to learn how to design integrated circuits. Unlike the study of high-level programming languages, which starts with the classic "hello world" program, it is unclear what the entry point should be for VLSI and what abstractions should be made along the way. Connecting courses across a curriculum to create a cohesive pathway to industry jobs and graduate school also presents a challenge. Additionally, the training materials for EDA tools (e.g., Cadence/Synopsys/Siemens Support, OpenROAD, EDAPlayground, NanoHUB) are ad hoc and suffer from accessibility, interoperability and versioning issues.

**Lack of trained educators**

Another key challenge is a shortage of adequately trained educators, even at the university level. Experienced faculty teaching VLSI are retiring and universities are finding it difficult to replace these faculty with recent university graduates. Those graduating with strong VLSI skills command high compensation from industry that most universities cannot match. Many universities are struggling to recruit faculty with the background and experience necessary to maintain let alone strengthen existing VLSI educational programs.

**Lack of trained support staff**

Supporting a strong VLSI program requires staff to maintain the systems used by the students. This support includes purchasing and maintaining workstations used to run the EDA tools, installing the EDA tools, installing/maintaining/debugging the PDKs, managing the licensing of the EDA tools, and performing regular updates to the PDK and tools. EDA tools requiring a single-seat license may be prohibitively difficult to implement at universities with minimal support staff, especially those unable to grant Virtual Private Network access to undergraduate students. At universities that cannot support dedicated staff, these tasks often fall on faculty members at universities that cannot support dedicated staff thus reducing their ability to focus on teaching and research.

**EDA/VLSI Costs and Challenges**

Teaching IC design effectively at the university level is challenging and includes a number of hidden costs that are easily overlooked. The EDA tools required to complete a successful IC design are very complex and involve experienced faculty and staff to install and maintain (see above). Although the EDA industry provides free or low cost licenses to universities for educational purposes, legal issues associated with these university licenses make it difficult for universities to share scripts, know how, and other relevant information to simplify their use and deployment in an educational setting. Open-source tools are increasingly being used as an alternative to proprietary tools but have more bugs and less support than proprietary tools. In addition, an IC tape-out as an educational activity is expensive and takes a long time. Besides the EDA tools, hardware lab resources and technical support are needed for students to test their chips. It is difficult to schedule sufficient time in an educational experience in which a student designs, tapes out, and tests a custom integrated circuit since most VLSI education has devolved into a single semester elective course. This particularly harms the ability of HBCUs, HSIs, Women's Colleges, and Tribal Colleges and Universities to provide a robust VLSI education.

**Limited Interest in IC Design**

Although IC design is an essential discipline supporting innovation in all areas of life, there is less interest among students in IC design than other more appealing emerging technologies. IC design appears to be viewed by students as a "back end" infrastructure activity with limited innovation, opportunity, and challenge. Bright students seem to be drawn more to Machine Learning (ML), cyber security, cloud computing, and other popular topics despite these areas having a strong reliance on hardware. Proper understanding of IC design, its opportunities, and technical challenges is limited among prospective students, parents, and counselors.

**Lack of Diversity in VLSI Design**

The challenge of creating a diverse workforce spans across all areas of STEM, but the problem is especially challenging for the VLSI field because: (1) the startup cost and resources needed to sustain a program in VLSI are often only available to R1 institutions and (2) very few Minority Serving Institutions have an R1 designation (including no HBCUs). Additionally, gender parity is lacking, though it is unclear how the field of VLSI/EDA compares to other ECE disciplines. There are broader efforts to attract women in engineering and computer science (e.g., Grace Hopper Conference, WISE, IEEE WIE) but less efforts for VLSI/EDA (e.g., Cadence Women in Technology Fellowship, DAC Women in Electronic Design Automation Achievement Award).

# Recommendations

**Mitigate the competing interests of IC Design**

Our overriding recommendation is to make IC design research, training, and workforce development a clear and visible focus of NSF and CISE in particular. There are competing interests for everyone's attention, and having NSF and CISE identify and support IC design research, education, and training, both as a discipline on its own and as a core component to interdisciplinary education, is critical to making an impact.

Our first recommendation is for NSF to support the development of bootcamps, tutorials, and educational resources for K-12 and university programs. The goal for the K-12 outreach is for educators to be able to excite their students about the world of IC design. This includes explaining how all their software apps and games need hardware to run them and the role of Moore's law in the advancement of technology in the world, including the latest iPads and bluetooth devices. The role of hardware should also be tied into existing programming languages and robotics programs.  At the undergraduate level, NSF should encourage the development of new programs in VLSI at institutions that have none, expand programs at existing institutions, and encourage the development of programs that highlight the interdisciplinary value of IC design by integrating robotics, biomedical engineering, or aerospace.  Many other fields rely on chips, but use only off-the-shelf hardware with limited applications. Emphasizing "what-if" questions can show students how advanced chips can lead to lower power, higher resolution, and new capabilities that spur innovation across fields. At the graduate level, revived support for graduate research in IC design is necessary to address the aging workforce in these areas. Moreover, continued challenges in emerging areas such as energy-efficient, sustainable, and secure computing remain a rich area of important research.

Our second recommendation is for NSF to support the development of public relations and media materials highlighting the importance of VLSI and ICs in society.  This could be as a suggestion that these materials would be welcome as a contribution to the broader impacts of CISE proposals in this area as well as creating a clearing house to host these materials for others to leverage.

On a related note, we believe NSF should support the development of contests at all levels of students to encourage their involvement in the field. At the K-12 level, this could be a development similar to FIRST Robotics, iGEM, MIT App Inventor and IceStudio; it could take the form of "Girls who Fab" and can focus on key exciting applications such as medical, neural implants, or block chains. Continued support of existing contests, such as the  IEEE SSCS PICO Design Contest [35], ICCAD CAD Contest [36], DAC System Design Contest [37], ISPD Design Contest [38] is useful to highlight the continued importance of these skill sets as well as tutorials such as the DAC Young Student Fellows Program [39].  Finally, encouraging maker-like spaces in which IC designs (possibly using FPGAs to start with) are a component of an interdisciplinary team effort would be useful to emphasize the role IC design has in the broader scope of innovation. This could include emphasizing concepts that connect to the entire electrical and computer engineering curriculum, including mixed-signal applications, AI applications, cybersecurity applications, robotics/controls applications.

Another option is to propose a "ChipCorps" similar to CSGrad4US [40] but for IC design, which can be motivated by the desire for sustainability, for energy-efficiency, and to save the planet, as well as national interest and patriotism. Finally, NSF could encourage the study of bringing IC design concepts into the Core STEM curriculum. For example, Boolean logic and transistors as switches can easily be incorporated into the middle school mathematics "Innovations" curriculum.

**Support shared educational materials**

In order to remove the barriers of teaching and learning at all levels of education, our second recommendation is to establish a centralized institution that can maintain good quality educational materials and remove teaching barriers. This could be a part of the National Chip Design Center (NCDC) or another institution that can act as a clearinghouse to develop and disseminate training, seminars, lectures, plug and play labs, and grading rubrics to support faculty for teaching at universities, colleges, and even individual learners. The institution should provide the infrastructure and know how to operate the vast level of support that VLSI education and training requires, such as license delivery and management, managing proprietary information, testing, etc. The workshops and training should provide entry-level, intermediate, and advanced content to encourage continuous learning at any level, or help transition the learner into new fields. For example, students trained in FPGA design to develop VLSI design skills, and support tools used in both FPGA and VLSI flows. The curriculum should support tape-out or "pseudo tape-out" (to layout but without fabrication) so that undergraduate learning is "full stack" with the connection between theory and reality. Education should be further expanded to include essential topics such as design for test (DFT), functional verification, system on chip (SoC), and others, and even different research focuses (photonics, quantum, ML, etc). With a central institution, materials for best practices and class material can be easily shared, updated and adapted. NSF should provide VLSI research grants to contribute educational materials to the centralized organization to further advance its library of materials and improve its workshops.

The development of a good textbook would help guide related courses and programs. In addition to the Patt and Patel book on "Bits to C" [41] a new book that is "Bits to Chips" which involves a chip fabrication at the end could be useful. In addition, the development of a freshman-level course to get students excited about computer engineering and IC design and cover "Apps to Silicon" may be useful. NSF should thus explore a model of a course bringing in BME/CS/MechE/Material Science students, pairing them with computer engineering students early in their programs to build an IC design (or even an FPGA design). This could be along the same lines as K-12 efforts teaching a 14-year-old how to build a chip. In all these cases, this discussion should include not only the feasibility of the scope of these classes but also the key learning objectives of the effort.

**Educator training for VLSI programs**

To address the lack of trained educators and an aging workforce in the area of VLSI and IC design, NSF should support summer training programs. These programs should be hosted at R1 institutions across the country and be designed to support the training of faculty in regional schools. Faculty should learn the tools and design flows as well as how to access centralized resources such as EDA in the cloud and FPGA development boards.

NSF should also support NSF Career type awards for young faculty focused on VLSI, particularly at schools with no or limited VLSI programs to encourage these schools to hire faculty in this area.

**Lack of trained staff**

A National Chip Design Center should be formed as a centralized place to support IC design programs, including the support for EDA tools and teaching assistants (TAs) that can be assigned to multiple institutions. These NCDC TAs can help with installation, access, and configuration of the tools needed for IC design. They should also support staff that can create IP and infrastructure to help with both tape-in and tape-out classes.

**EDA/VLSI costs and challenges**

To address the costs and challenges associated with EDA tools and tape-out classes, NSF should continue to support open-source tools, design flows, best practices, sharing of scripts and other resources by creating a support community of educators. They should continue to involve EDA vendors to facilitate greater interchange of best practices. NSF should also fund the creation of centralized regional resources for testing and other capital equipment that can be used by multiple universities as a part of the NCDC. This testing facility should combine short in-person training w/ remote FPGA based testing. They can also encourage large universities to share computing resources and licenses with neighboring institutions.

**Diversity & Inclusion**

Research must be done to understand the baseline issues in teaching VLSI and IC design. Using these results, VLSI education should be extended to prepare more students for graduate VLSI training. For example, R2 students need preliminary training to be able to enter an R1 graduate program in VLSI.

To address the lack of gender and racial diversity in the VLSI field, the problem must be understood. Research studies should be conducted to understand where minoritized groups are enrolled and identify ways to ensure that their institutions have pathways into the field. This includes ensuring that every minority-serving institution has a sustainable VLSI program or a regional partner that provides full access to a VLSI program.

These studies should also address the appropriate learning pathways to adequately educate a learner from different socioeconomic backgrounds. How can we effectively address and influence students' awareness on the importance of VLSI training / chip design? What strategies are effective in attracting students into the semiconductor industry?

# EDA Tool Flows

The Electronic Design Automation (EDA) industry is relatively young and only gained popularity starting in the early 1980s. This was largely fostered by the increased number of computer engineering (as opposed to only electrical engineering) designers during the Mead-Conway era [42] as well as the advent of the Application-Specific Integrated Circuit (ASIC) design style using hardware description languages, synthesis and standard cells.

One of the earliest and most successful EDA tools was for circuit simulation as a Berkeley class project in 1969 that ultimately led to SPICE [43]. Similarly, the creation of numerous layout tools such as Magic (1984) [44] and Electric (1983) [45] helped foster the Mead-Conway designs using Scalable CMOS (SCMOS) design rules through the MOSIS fabrication service.

The first Place and Route (P&R) tools were created in the mid-1970s for ASIC designs and gained usage during the fabless semiconductor business model first offered by companies like VLSI Technologies (1979), LSI Logic (1981) and later TSMC (1987). These tools led to the first commercial EDA companies in 1981 and the founding of the Design Automation Conference in 1984. The EDA Consortium (EDAC) was founded in 1988 and was a long-time co-sponsor of DAC along with IEEE and ACM. Several companies such as IBM, Intel and TI have had a history of their own EDA tools for a Integrated Device Manufacturing (IDM) business model. Only Intel and, to a lesser extent, TI still do this.

Proprietary, commercial tools gained popularity in academia in the late-80s and 90s due to their active development, high performance, and cheap (or free) academic licenses. Entire conferences were created by educators with publications on best practices using proprietary EDA tools [46]–[48]. In particular, the IEEE International Conference on Microelectronics Systems Education (MSE) was held in odd years from 1999-2018 while a "sister" conference, the European Workshop on Microelectronics Education (EWME), was held in the even years. These conferences have largely devolved with the decline in VLSI education and exist, at the best, as a single track in larger conferences.

As of today, EDA is an approximately $13 billion industry of tools for design, verification, and testing of semiconductor chips. The "big three" EDA companies continue to be Synopsys, Cadence, and Mentor Graphics (now a part of Siemens) and cover most of the market. While there are occasional successful start-ups such as Magma Design Automation, Avant! Corporation, Cascade Design Automation, and Apache Design Systems, to name a few, the most common outcome has been acquisition by the big three.

Going forward, it is important to understand the increasingly central role of EDA technology in IC design. Systems innovation depends on having a capability to perform design; the well-lamented *design capability gap* means that designers are less and less able to use the availability transistors on a chip at a leading-edge node. Design capability in turn depends on design *automation* technology, i.e., EDA. In his recent opening keynote at the 2022 Design

Automation Conference, Mark Papermaster (AMD CTO) pointed out that more than half of the benefit seen at the 3nm node comes from design-technology co-optimization (DTCO), and that this proportion is increasing rapidly. Via DTCO and continued algorithmic and functional innovations, EDA enables the "design-based equivalent scaling" which has been a key element of the Moore's law value trajectory in this century. Thus, NSF should support education, research and workforce development not only for VLSI design, but for the underlying technology of EDA as well.

## Summary of Current State

EDA companies typically provide reference flows for their proprietary tools, but these flows do not cover the entire design process which usually requires tools from multiple vendors and specific settings for different technologies and their associated PDKs. Because of this, semiconductor companies and academic institutions must develop their own design flows. As a result, there are a huge number and variety of flows with little consistency. The cost of developing, debugging and supporting these flows far exceeds the resources of most academic institutions.

The dependence on cheap and free academic licenses from proprietary vendors may seem generous, but in fact has a negative impact on research independence and willingness to work on interoperability. In particular, most licenses specifically disallow benchmarking comparisons which stymies research. On occasion, academic institutions have attempted to modify and share proprietary tools flows, but companies have threatened to revoke tool licenses because they do not want information getting in the hands of competitors. If academics lose access to tools required for their research by doing something which impacts a company, that area of research will be fundamentally off limits.

Proprietary tools come with virtually no support. Usually one or two designees can access online help and all questions from instructors and students must go through this person. Usually at least one designee is an IT staff member with no design knowledge. Open-source tools, on the other hand, are entirely community supported so the quality and level of support is highly sporadic. Documentation of open-source tools is often lacking or outdated.

Recently, some courses have begun using the OpenROAD [46], OpenLANE [47], and the Caravel/OpenMPW [51] infrastructure from Efabless using the SkyWater 130nm open-source PDK [26]. While these are highly accessible compared to the proprietary tools, they suffer from support issues and the tools are mostly volunteer supported so they have stability issues. In addition, while they teach TCL/Python scripting, baseline flows, understanding of analysis reports and other basic skills, they are ultimately different from what is used in most industry jobs today.

# Key Challenges

### Access barriers

For workforce development, it is important to remove barriers to access tools, flows and design enablements so that a new generation can be excited and attracted by VLSI design, as early in life as possible.

### Standards and interoperability

Today's design flows rely on standards that are not actually open. For example, a standard format to support physical verification (e.g., a "design rule check") does not exist and there is no open format for system design constraints. A stable foundation for research and development, demands fully open source reference implementations and libraries. Funding policies for research that require open-source reference implementations and libraries are absent today.

### Reproducibility

The advancement of science and knowledge depends on reproducibility. Forward progress is impeded when methods used to reproduce reported results cannot be freely and openly shared. Onerous and restrictive provisions are imposed by proprietary tool vendors, and this has a chilling effect on progress in the IC design and EDA realm. Reproducibility and collaboration between academia and industry often requires complicated legal agreements. NSF is not following its own policy in the areas of IC design and EDA research about the publicly accessible artifacts in data management plans [52]. In particular, according to Code of Federal Regulations (2 CFR 215.36), researchers are required to share research data which is defined as "the recorded factual material commonly accepted in the scientific community as necessary to validate research findings." This should include EDA tools and flows to reproduce results.

In the case of open-source EDA, this impediment does not exist but comparisons are often not considered "state of the art" for publication by many reviewers.

### Applicable training

For job readiness, training on the latest commercial tools used by industry is highly desirable. It should be noted that these "commercial" tools could be open-source tools or could be proprietary tools, but most of the industry currently uses proprietary tools. It is desirable that any training be done on the same tools to reduce the amount of on-the-job training. However, some essential skills are agnostic to the specific tools a student may have used and instead rely on concepts that the broader set of tools utilize.

**Maintenance costs**

Design flows and open-source tools require maintenance (e.g., training materials, tool updates, evolving computing platforms, and changing libraries), which is an ongoing cost.  The cost of this maintenance can be minimized, but is very difficult to eliminate entirely. Research projects that release open-source projects are often abandoned after critical student contributors graduate. The NSF and other research agencies do not traditionally fund ongoing maintenance.

## Recommendations

**Improve accessibility**

The NSF should actively support mechanisms and infrastructure for design flows that are able to produce industry quality results for training, exploration and research. This will reduce waste and redundancy of efforts such as tool bring-up, tuning, maintenance; designer and education/course support; development and maintenance of educational/lab materials; etc. One or more "canonical flows" should be defined and maintained for both proprietary EDA tools and open-source tools should be available in advanced nodes, older nodes and with FPGA targets.

We recommend that the NSF establish a service, possibly at a National Chip Design Center, to centrally maintain these flows and provide a quality assured easy "bring-up". For example, CMC Microsystems (Canada) provides "design methodologies [to] help guide users through the design process to ensure successful designs" [53].  This includes:
> **Design Flows**—Step by step instructions guiding users through the CMC design environment on knowing "what to do"
> **Design checklists**—a list of items to consider in your design to ensure successful implementation
> **Design tutorials**—instructional material to go through the design flows. It may also include sample design files and libraries." [53]

Similarly, Industrial Technology Research Institute (ITRI, Taiwan) and IMEC/EUROPRACTICE (Europe) have provided examples of structured flows and hand-off points that have dramatically lowered the barriers for academics.

**Promote sharing of methodologies and flows**

The NSF should fund research projects that share methodologies and flows. This should not only include initial releases but ongoing maintenance and support for the community. Success of projects should extend beyond research publications and citations, and instead include the number of users. Similar to the broader impacts required of research topics, there should be a "community support" component of all proposals.

An example of how a fully open source flow that is shared between academic and industrial use cases is the area of machine learning. When Google open sourced the TensorFlow environment, the amount of ML research and development not only significantly accelerated with TensorFlow but also alternative solutions like PyTorch dramatically increased as well. Similarly, other fully open source reference flows for software like LLVM and the GCC toolchain suite have helped bridge the gap between academic research and industry. In response, companies like Google, Microsoft, Amazon, Intel, and AMD have provided a huge number of software engineering resources to these projects which allow researchers to continue to focus on new innovation.

**Promote common formats for data interchange**

We recommend that NSF support efforts to (i) identify current gaps and risks in flow development and distribution; (ii) define open canonical formats for data interchange; (iii) develop reference implementations; and (iv) support the adoption of these formats.

With respect to standards and intermediate formats, having a policy similar to W3C's requirement of multiple independent implementations with at least one fully open solution before adoption will help create a stable and lasting foundation for research and development.

An important benefit of having common formats is to promote the previous goals of accessibility and sharing. This is seen in "flow runner" platforms (e.g., Hammer, mflowgen, SiliconCompiler, OpenLane), which enable interchangeability of tools (e.g., logic synthesis or place-and-route) between chip design states (e.g., synthesizable RTL, or floorplan DEF).

**Role of open-source EDA tools and flows**

The simplest first step for establishing these canonical flows and sharing them would be to support the development and maintenance of a flow for open-source tools and open enablements (i.e., PDKs and libraries).  This is a special and highly impactful combination that maximizes simplicity by avoiding the need for NDAs, export control, and other potential barriers to access. The ultimate goal for this service would be to maintain a flow (or family of flows) that is tool agnostic, allowing a variety of open-source and proprietary tools to be used within the flow.  OpenROAD and OpenLANE are a good starting point for this flow as they allow users to tweak different components of the tool itself.  OpenROAD is the core of the IEEE CEDA Design Automation Technical Committee's "Robust Design Flow" (RDF) from 2019 and has seen impressive growth, but its future is uncertain. There are also many missing components.  A fully open-source, editable, flexible and extensible RTL-to-GDS tool chain is something that proprietary IC tools cannot provide. Moreover, open-source tools serve several fundamental roles that proprietary tools cannot [54]. (i) Open source enables EDA workforce development by teaching EDA algorithms, software architecture and software development through source code itself. At a DAC-2022 panel, Dr. Charles Alpert of Cadence noted that development experience in OpenROAD can save two years of learning curve once the developer is hired into his organization. (Insight into EDA tools also helps develop innovators in design methodology, as opposed to "button-pushers".)  (ii) Use of open-source tools can be freely scaled to enable system-level, algorithm and architecture explorations that are infeasible today with proprietary tools. (iii) Open-source tools can generate data that is freely sharable; importantly, this unblocks research in AI and machine learning for IC design.

These canonical flows are not meant to be a direct competition with proprietary tools. However, some of the developments/innovations that come out of OpenROAD may be integrated/adopted by commercial vendors. Companies will not invest in tooling which they are unable to use and they will only invest minimally if the cost of moving research into production is exceptionally high. For example, from Google's philosophy on research page [55]: "We have always seen scientific publications as an important component for much of our research work, but for fundamental research projects, open-source code releases and new datasets can be particularly valuable."

NSF should encourage research on new flow abstractions and EDA tools. Hammer, mflowgen, SiliconCompiler, and OpenLane can all be viewed as providing easy-to-understand, retargetable flow abstractions. Different tools can be plugged in to flow steps (e.g., synthesis, or detailed routing) including open-source or proprietary. While the Mead-Conway era raised abstractions to digital design, we need a new era that raises abstractions to enable software developers to design the next generation chips.

# Intellectual Property (IP)

Intellectual Property (IP) refers to semiconductor design components that can be reused. Usually these are utilized in a System on a Chip (SoC) in the form of digital, mixed-signal or analog cores and can even be the processing core itself, a memory interface, a co-processor, a network interface, a sensor, etc. Besides cores, standard cell libraries, input/output (IO) pad libraries, and memories (compilers, controllers, etc.) are considered IP.

IP reuse allows designers to use pre-made designs and increase overall productivity. IP can be provided as either hard (layout level for a specific technology) or soft (behavioral level and portable among technologies through synthesis, placement and routing to create the layout) depending on the requirements. Hard IP is frequently preferred for analog and mixed-signal whereas digital IP is often behavioral and implemented in a Hardware Description Language (HDL) such as Verilog.

Design verification is probably the most time intensive component of design and IP reuse allows designers to assume that cores are verified and reduce the overall design and verification effort. However, the interface with design IP may require additional verification and SoC designers, therefore, focusing on system verification. Entire companies, such as ARM, have made businesses out of licensing IP to customers and either charging license fees or royalties on resulting designs.
Companies either license IP from third parties or have their own set of IP. Because of this, there are multiple layers of secrecy and underlying infrastructure, all of which make it difficult for academics to reproduce designs or teach advanced topics involving entire systems.

## Summary of Current State

There is a large gap between what education and research institutions have access to and what is needed to make a design viable. Some R1 institutions have access to IP through industrial partnerships, but the majority have limited or no access.

**Cell Libraries**

Proprietary, commercial libraries are challenging to obtain. Similar to the discussion in the technology chapter, there are numerous issues with complex licensing and data sharing. Often, "black box" libraries and models are provided that do not have the entire layout and instead just have "abstract" views of the cells (or cores). This doesn't allow students to see internals and prevents full-stack exploration. These libraries are available from fabs (e.g., TSMC, Intel) and third-party providers (e.g., ARM).  In addition,  some companies may prevent certain cell libraries from being utilized for research and/or instruction due to contract verbiage.  This ultimately hurts edifying students with the knowledge on how to use standard-cell libraries properly.

**Memories**

Memories are a fundamental building block such that they are as important as cells. Memories can be either available as hard IP blocks of specific sizes or they can sometimes be generated by a memory compiler. Many memory compilers have limited ranges of sizes and options for memories that they can produce. There are also numerous memory variants besides standard Static Random Access Memories (SRAMs) such as register files, FIFOs, etc.

Proprietary memory compilers may be provided by a fab or a third-party, but they often are immutable and may provide only black box views of the memory itself by omitting all internals until a fabrication agreement is signed. Memory compilers are available with agreements through TSMC or other fabs.

There is one open-source, actively developed memory compiler called OpenRAM [56], [57]. This has been used to produce silicon verified memories in SkyWater 130nm and it also supports FreePDK45 and AMI 0.35/SCMOS. OpenRAM has been in partnership with the Google/Efabless OpenMPW program and has had cores on every MPW design as a part of the Caravel test harness and on many of the user projects. There have been 5 OpenRAM test chips fabricated and the first has been received and verified as functional.

**Soft/Hard Digital/Analog Cores**

The most notorious specification is the RISC-V Instruction Set Architecture (ISA) which has received broad enthusiasm and adoption [18], [58]. There have been many RISC-V cores made available [59]–[63].

There are a few open-source IP repositories, but they are not very active, in general. For example, OpenCores.org has a large number of cores with many of them being unverified . There was a repository called UMIPS that had both digital and analog cores but has not been updated in nearly two decades [64]. Moreover, the IP was mostly for a single 180nm process from TSMC and required an NDA.

There are too many companies with proprietary IP to list, but these include ARM, Dolphin, SiFive, Synopsys, etc. These all provide high-quality IP but require licenses and may provide only "black box" models until tape-outs are planned.

**Standards**

There have been a few efforts to standardize how IP is distributed. These include database models as well as structures and tools.
There was a push towards using an "open" database standard called OpenAccess, but this is not actually an open source solution and requires complex agreements and membership to gain full access. These formats are not standardized nor do they have a format that everyone uses. This ultimately leads to problems that relate to how IP can be utilized effectively and efficiently.

IP-XACT is an XML format that provides a standard structure for sharing IP [65]. FuseSoC is a package management system for distributing and integrating IP [66].

There are a large number of interface standards that also support IP reuse. These include the open-source Wishbone bus [67] and commercial buses such as ARM AMBA [68], IBM CoreConnect [69], and many others.

## Key Challenges

The requirements to build a modern IC requires a lot of supporting infrastructure which is mostly inaccessible to academics. In recent decades, standard cell libraries became accessible in a limited fashion for either older technologies or through restrictive licensing. Recently, memory compilers, including regular SRAM and non-volatile memory, remain a considerable challenge.

One of the fundamental challenges with IP, however, remains verification and documentation. Verifying IP is essential to making useful and reliable IP that anyone can use.

**Fundamental IP**

There's a lack of access to IP that is high quality, easy to use, and fundamental and easily modifiable and customized. This lack has a significant impact on reproducibility of research results and training of students. At a minimum, complete digital systems are needed such as:
- Standard cell libraries
- Auxiliary cells (taps, level shifters, etc.)
- I/O pad cells
- DFT-enabled cells
- Memory compilers (Flash, SRAM, DRAM, RRAM)

In addition to the above minimum set of IP, there are additional mixed-signal and analog components that are needed for any system that interacts with the physical world including:
- Phase Locked Loops (PLL)
- Analog to Digital Converters (ADC)
- Digital to Analog Converters (DAC)
- Universal Serial Bus (USB)
- Voltage Regulators (e.g., LDO)

**Complex and restrictive licenses**

In order to access IP for education, research and prototyping, complex NDAs are required in addition to possible access fees or even fabrication agreements. It is not possible for most universities to sign such agreements which restricts access to a select few R1 schools with industry connections.

The proliferation of "academic only" and "non-commercial" licensing methodologies for the release of research assets (in all of EDA flows, IP and other areas) prevents collaboration between academia and industry, reducing the incentive for industry to help provide investment. This sometimes causes issues with how libraries are licensed and may lead to not being able to be used in the classroom. Ultimately, students tend to suffer as they typically cannot get access nor can they get access to full-featured and commercial grade standard-cell libraries. Sadly, this leads to students being not trained in the area that is common in industry.

**Lack of maintenance and verification**

The open-source IP that is available is poorly verified and not maintained. Many research projects do not share resulting IP at all. Those projects that are made available typically remain static after initial creation since students graduate and there are no resources for bug fixes, enhancements, and other improvements. In general, IP that is created is verified "just enough" to prove a research project and further work only delays graduation and does not lead to faculty advancement. EDA tool flow upgrades may require changes to IP, new features may need to be added or expanded, and new use cases might need to be supported.

## Recommendations

The recent success of the RISC-V ISA has shown that open-source is a viable solution for many things. In addition, open-source can enable commercial use and does not require proprietary solutions for successful business models. These recommendations center around the development and support of high-quality, open, sharable IP and encouraging its use.

**Support IP development**

The creation of high quality, fully open source IP for fundamental areas like memory and connectivity need significant investment. These IPs need to be accessible to everyone, work across FPGA and ASIC implementations and be compatible with reference EDA tool flows as previously recommended. Although FPGA and ASIC implementations are different, they are both important. For example, FPGA implementations can be utilized for validating architectures and designs before they go to fabrication for an ASIC implementation. However, both FPGA and ASIC have different mechanisms for design flows. Consequently, there needs to be better support for IP development for both areas that can help enhance education as well as promote better research ideas.

NSF should invest in projects that generate IP that impact not only technology but also workforce development. The success of these projects should include a component of broader impact through the number of users and usage in research projects and publication.

A centralized source of IP, such as a National Chip Design Center, would enable tracking of user success and integration. This would benefit others by connecting researchers who have utilized a certain IP to address problems. The central resource would lead to more users and faster adoption which would ultimately improve the quality of IP.

NSF should fund strategic "grand challenges" to address new fundamental design problems that are essential to users. This could be through specific competitions to help create grand challenges that researchers need to solve. This is similar to the NSF MRI awards but specifically for semiconductor IP.

**Promote maintenance and verification in addition to development**

NSF should make the maintenance of IP as an essential component of Data Management Plans. While NSF proposals need to have broader impact and intellectual merit, there should be a responsibility to continue disseminating results. In addition, NSF should create mechanisms for follow-on funding to sustain successful IP produced by research proposals.

To ensure scientific rigor, research projects should be required to be reproducible by the community and a plan to continue that reproducibility. Proposals should have a section devoted to correctness and verification to prove that results are valid through simulation, formal methods, or other ways. These same methods can then be used to maintain IP as EDA tools change, technologies scale, and other assumptions are changed. Traditionally, this has led to "stale" IP that is no longer useful.

**Centralized infrastructure and distribution**

Systematic mechanisms should be available for building, verifying and distributing IP. For example, Python has good methods for helping distribute IP (e.g., pip install, npm, etc.). There are some approaches such as FuseSoC [66], but this is not a standard. A National chip Design Center should host centralized repositories and have recommended continuous integration policies for IP generated by NSF funded projects.

NSF should support open-source verification test suites for critical IP blocks. These verification test suites need to be evaluated by their adoption by IP block developers and enablement of new IP blocks to be developed quickly. More importantly, since these blocks are critical to many architectures, it is important that the availability is paramount to US researchers, academics and commercial institutions.

IP blocks need to ensure compatibility with reference design flows
- Compatibility / support with reference EDA flows
- Integration with CI for EDA reference flows

**Require reproducibility of research**

NSF should require reproducibility of research projects. That is, NSF should have a panel who is responsible to ensure that any grants that are selected are reproducible. Also, this group should be responsible for seeing how well a certain work force can deploy these ideas to create effective ideas. NSF should have a mechanism that data that is produced and can be reproduced. Reproducible enables access and high quality and can help communicate information to industry that what is being developed can be beneficial.

**Industry collaboration**

Promoting industry collaboration is important for future innovation. This is predicated on the idea that open-source IP is incredibly important for the future of scientific discovery. Researchers need to try ideas and this open-source IP can help facilitate more streamlined and efficient architectures and systems. Open-source is an incredibly important piece of the puzzle allowing users to try different ideas and, perhaps, solve long-standing problems that are previously intractable. In addition, having open-source hardware can facilitate collaboration between researchers as well as the possibility in creating repeatable instances and infrastructure. This could be a requirement to include a plan to help create sustainability to allow future funding.

Can NSF support companies to help foster interaction with research institutions? Companies may be able to promote joint efforts with academic institutions if companies can partner with academic institutions to effectively achieve solutions that help many. Industry collaboration may also help foster interaction and ideas for improvement that academic institutions can integrate and implement. This leads to better ideas to address scientific discovery, issues, and challenges that may stymie one branch of an industry.

**Documentation and training**

Documentation and training is also important for helping take IP and using it effectively. Many institutions do not have access to information to help integrate designs. Having educational initiatives that can help training of individuals and how to use that IP effectively. This could also be integral to future workers in the United States by promoting the understanding of what semiconductors are and how to use them effectively.

# Packaging and Testing

The traditional functions of electronic packaging are to support signal connections, deliver power, remove heat and protect the embedded chips from mechanical and environmental threats. Advanced packaging offers value-added functions beyond these, including heterogeneous integration of devices based on a range of materials, chipletization of large die into a set of smaller die in optimized processors, and opportunities to dramatically increase the bandwidth across the system. For high end digital systems, the future is likely to focus on large scale use of interposers to integrate multiple 2D and 3D (stacked) chips into a module, and the use of photonics to communicate signals between modules. More systems are being built using fan-out and fan-in wafer level packaging (FOWLP and FIWLP) to enable space and cost efficient integration. Mixed signal and power delivery applications are going to benefit from heterogeneous integration of III-V and silicon devices, for example GaN high voltage and power devices, or InP high frequency devices integrated with high density CMOS technologies. Many applications will benefit from a CMOS+X approach where devices based on a non-silicon technology are fabricated on top of a CMOS wafer. A common example is the integration of resistive RAM (ReRAM) on top of CMOS to enable non-volatile storage and/or processing in memory.

## Summary of Current State

At present, 90% of packaging manufacturing is conducted in the Asia Pacific region. Research access to advanced packaging is either non-existent or difficult and expensive to arrange. The typical cost of an IC in 5nm technology node is over $500M with 50% of that attributed to the fabrication and packaging.

The packaging landscape is also changing rapidly. Interposers are taking over high end packaging today and photonics will be taking over from electrical interconnect in inter module communications. Heterogeneous integration will create paradigm shifts in analog and power electronics, with the intimate integration of silicon, III-Vs and other material sets. Recent process technologies of fan-out wafer level processing (FOWLP) and fan-in wafer level (FIWLP) require IC design and packaging executed concurrently. However, interposers can be built using legacy technologies, such as provided by Xfab, SkyWater, and nHanced Semiconductors. The latter two also support a range of integration technologies, including bumping, hybrid bonding, etc.

The codesign notion of IC design and packaging requires design aids for engineers. For IC design, the foundry-provided PDKs are the tools that engineers use to complete the IC design. Something equivalent is needed for packaging called an Assembly Design Kits (ADK) with ADKs being parallel to PDKs. ADKs contain a combination of EDA and Outsourced Semiconductor Assembly and Test (OSAT) capabilities.

Only a handful of universities in the CISE community offer courses in packaging. Fewer yet offer coordinated course sequences. Few VLSI courses include any material on packaging and related developments. There is no systematic curricula at institutions of higher learning regarding packaging, especially, design of packages so that ICs can be tested.

Custom packaging for chips produced in university research projects is often difficult to source, especially for modern commercial quality packages.  It's common for universities to use off-the shelf packages or chip-on-board integration.

Testing of chips produced in an educational setting can be managed with specialized test boards and PC based instrumentation, but it is done in an ad hoc manner and each group reproduces a similar infrastructure.  In contrast, the test and debug of chips produced in a research setting requires considerable equipment support, such as probe systems, logic analyzers, etc. More advanced technologies require elaborate laboratories to test these high density ICs.   Not all universities have these capabilities with the problem being more evident in lower tier schools.

## Key Challenges

### University access to advanced packaging

University access to advanced packaging technologies today is very limited.  As a result a jury-rig assembly approach is often used.  For example, many projects are packaged by wire-bonding a chip to a pre-existing off-the-shelf package or a Printed Circuit Board and using micro-probes to introduce or measure high speed signals.  (Wire bonds don't support high frequency signals.)  While sufficient for verifying a circuit function it does little towards training the PhD students in system design and co-optimization.  If a university researcher wants to use advanced packaging, they have to organize their own manufacturing flow from the very limited range of vendors willing to build R&D lots.  For example, a university researcher might arrange for a foundry to process MPW wafers only through to a top metal and divert the wafers to a speciality fab for incorporation of CMOS+X or Heterogeneous Integration steps.  This is expensive and time consuming.  A better solution is needed.

### Lack of US based packaging industry

90% of packaging manufacturing is conducted in the Asia Pacific region.  There is a need for talent and technological knowledge to create opportunities for industry onshoring.  Few universities offer a packaging course, let alone a packaging set of courses. Few microelectronics courses include a module on advanced packaging.

### Chip/package co-design

Optimal codesign of the packaging and the embedded chips present a tremendous opportunity for improved size, weight, performance, power and cost.  The co-design notion of IC design and packaging executed concurrently requires design aids for engineers.  For IC design, this is the foundry-provided PDKs are the tools that engineers use to complete the IC design. Something equivalently needed for packaging called Assembly Design Kits (ADK). ADKs parallel PDKs. They contain a combination of EDA and Outsourced Semiconductor Assembly and Test (OSAT) capabilities. The co-design of PDKs/ADKs can reduce fabrication/package cost by at least a factor of 75%.

**Many universities lack facilities to test research chips**

Testing is needed to verify the outcomes of a university project whether it is fabricated for research purposes or educational ones. Testing of class tape-outs can be facilitated by incorporating test interfaces on a chip along with predesigned boards and the use of PC-based test equipment. However, the testing of research systems are more complex, requiring access to advanced test equipment (i.e. probe stations, microscopes) and the know-how to use it. Many universities lack such access.

# Recommendations

**NSF should establish a Call for Packaging Research Infrastructure**

That NSF should establish a Semiconductor Packaging Research Infrastructure program, or include it in a broader program. This should provide organized, systematic access to a suitable range of advanced packaging and integration technologies to enable prototyping. This call should encourage research activity to understand the ADK optimal content and the co-design interaction between ADK, PDK, and EDA tools. Ultimately, open-source ADKs should be made available and recommended flows of both proprietary and open-source tools should support these ADKs.

**NSF should establish a Call for Chip/Package codesign and design with Heterogeneous Integration technologies**

The NSF should establish a program specific to exploring the opportunities that arise from CMOS+X, advanced packaging, and heterogeneous integration technologies. The program should leverage the infrastructure that is implemented in the program described above and should encourage projects in Design Technology Co-Optimization (DTCO), chip-package co-design tools and flows, optimal heterogeneous integration, and the design, fabrication and testing of case studies. Packaging on chiplets on an interposer provides unique research opportunities on interfaces - there is a need to establish a community working towards some standards

**NSF should establish a Call for Shared Test Infrastructure**

A National Chip Design Center (NCDC) (or regional labs at specific universities) should provide test equipment and infrastructure along with their support and training. While it may not make sense for individual schools to have this equipment, it makes sense for researchers and students to travel to gain access to it. The call should establish a community of stakeholders (e.g., either nationally or regionally) that can share semiconductor instrumentation equipment for testing. There should be some capital equipment awards for universities to establish these testing facilities and ongoing support to provide maintenance and support for other researchers. These facilities should be readily available with minimal or no fees.  The community should also be encouraged to conduct sharing of key IP to aid testing including IO interfaces, designs, etc.

# Conclusions

This workshop report presented the key challenges and recommendations in the most-important areas identified by its participants, namely: Technology Nodes, Training and Education, Electronic Design Automation (EDA) Tool Flows, Intellectual Property (IP), and Packaging and Testing. All participants agreed that there are severe challenges in each area that need immediate attention. The two most prevalent solutions appear to be embracing both open-source and proprietary supported IP, design kits, and tool flows as well as a national plan to coordinate this. The establishment of a National Chip Design Center (NCDC) similar to other regions would be the most effective way to create, maintain, and provide access to the broader US VLSI design community.

# Appendix A: Virtual Workshop Attendees

| Name | Association |
|---|---|
| Timothy Ansell | Google |
| Krste Asanovic | UC Berkeley |
| Iris Bahar | Brown University |
| Mamta Bansal | Qualcomm |
| Christopher Batten | Cornell University |
| Peter Beerel | University of Southern California |
| Sanjukta Bhanja | University of South Florida |
| Nitin Borkar | Qualcomm |
| Erik Brunvand | National Science Foundation |
| Ben Calhoun | University of Virginia |
| John Damoulakis | Information Sciences Institute |
| Donald Davidson | Synopsys |
| Rhett Davis | North Carolina State University |
| Antonio de la Serna | DARPA |
| Jeff DiCorpo | Efabless |
| Tim Edwards | Efabless |
| Paul Franzon | North Carolina State University |
| Hui Fu | Intel |
| Pierre-Emmanuel Gaillardon | University of Utah |
| Matthew Guthaus | UC Santa Cruz |
| Deirdre Hanford | Synopsys |
| David Harris | Harvey Mudd |
| Patrick Haspel | Synopsys |
| LaMar Hill | NY Creates |
| Tina Hudson | Rose-Hulman Institute of Technology |
| Angela Hwang | Synopsys |
| Bill Isaacson | Muse Semiconductor |
| Riadul Islam | University of Maryland Baltimore County |
| Mark Johnson | Purdue University |
| Siddharth Joshi | Notre Dame |
| David Junkin | Cadence |
| Andrew Kahng | UC San Diego |
| Mohamed Kassem | Efabless |
| Srinivas Katkoori | USF |
| Brucek Khailany | NVidia |
| Allan Klinck | Siemens |
| Anton Klotz | Cadence |
| Serge Leef | DARPA |
| Daniel Limbrick | North Carolina Agricultural and Technical State University (NC A&T) |
| Ken Mai | Carnegie Mellon University |
| Rob Mains | CHIPS Alliance |
| Rajit Manohar | Yale University |
| Pinaki Mazumder | National Science Foundation |
| Gayatri Mehta | University of North Texas |
| Ross Miller | SkyWater |
| Robert Montoye | IBM |
| Matthew Morrison | Notre Dame |
| Anna Nelson | Cohen Group |
| John Nestor | Lafayette College |
| Borivoje Nikolic | UC Berkeley |
| Adrian Nunez-Rocha | Qualcomm |
| Kenneth O | University of Texas Dallas |
| Christopher Ober | Cornell University |
| Andreas Olofsson | Zero ASIC |
| Ron Olson | Cornell University |
| Geoff Porter | Muse Semiconductor |
| Jordan Roth | Cohen Group |
| Mehdi Saligane | University of Michigan |
| Sophia Shao | UC Berkeley |
| Mircea Stan | University of Virginia |
| James Stine | Oklahoma State University |
| Christoph Studer | ETH Zurich |
| George Suarez | NASA |
| Michael Taylor | University of Washington |

| Jim Wieser | Texas Instruments |
| Michael Wishart | Efabless |
| Qing Wu | Air Force Research Labs |
| Todd Younkin | Semiconductor Research Corporation |
| Victor Zhirnov | Semiconductor Research Corporation |
| Zoran Zvonar | Analog Devices |

# Appendix B: Virtual Workshop Agenda

| Time | Speaker | Association |
|---:|---|---|
| **October 14, 2021** | | |
| 9:00 | Erik Brunvand | National Science Foundation |
| 9:20 | Todd Younkin | Semiconductor Research Corporation |
| 9:40 | Qing Wu | Air Force Research Labs |
| 10:00 | George Suarez | NASA |
| 10:20 | **Government Q&A** | |
| 10:40 | Ken Mai | Carnegie Mellon University |
| 11:00 | Gayatri Mehta | University of North Texas |
| 11:20 | Daniel Limbrick | North Carolina Agricultural and Technical State University (NC A&T) |
| 11:40 | Tina Hudson | Rose-Hulman Institute of Technology |
| 12:00 | Kenneth O | University of Texas Dallas |
| 12:20 | Michael Taylor | University of Washington |
| 12:40 | **Academia Q&A** | |
| | | |
| **October 15, 2021** | | |
| 9:00 | Geoff Porter | Muse Semiconductor |
| 9:20 | Hui Fu | Intel Corporation |
| 9:40 | LaMar Hill | NY CREATES |
| 10:00 | Christoph Studer | ETH Zurich |
| 10:20 | Ross Miller | SkyWater Technology |
| 10:40 | **Foundry Q&A** | |
| 11:00 | Andrew Kahng | UCSD/OpenRoad/Startups |
| 11:20 | Rob Mains | CHIPS Alliance |
| 11:40 | Tim Ansell | Google |
| 12:00 | Mohamed Kassem | Efabless |
| 12:20 | Zoran Zvonar | Analog Devices |
| 12:40 | **Industry Q&A** | |

# Appendix C: In-Person Workshop Invitees

| Last | First | Affiliation |
|---|---|---|
| Batten | Christopher | Cornell University |
| Brunvand | Erik | NSF/University of Utah |
| Gaillardon | Pierre-Emmanuel | University of Utah |
| Guthaus | Matthew | University of California, Santa Cruz |
| Manohar | Rajit | Yale University |
| Mazumder | Pinaki | NSF/University of Michigan |
| Stine | James | Oklahoma State University |
| Harris | David | Harvey Mudd College |
| Junkin | David | Cadence |
| Damoulakis | John | Cadence |
| Jeeawoody | Shakeel | Mentor Graphics |
| Kassem | Mohamed | E-fabless |
| Mai | Ken | Carnegie Mellon University |
| Morrison | Matthew | University of Notre Dame |
| Bahar | Iris | Colorado School of Mines |
| Stan | Mircea | University of Virginia |
| O | Kenneth | University of Texas at Dallas |
| Taylor | Michael | University of Washington |
| Limbrick | Daniel | North Carolina Agricultural and Technical State University |
| Davis | Rhett | North Carolina State University |
| Kahng | Andrew | University of California, San Diego |
| Franzon | Paul | North Carolina State University |
| Wirthlin | Mike | Brigham Young University |
| Murmann | Boris | Stanford University |
| Beerel | Peter | University of Southern California |
| Chang | Lifu | MOSIS |
| Muldavin | Jeremy | GlobalFoundries |
| Fu | Hui | Intel |
| Chetan | Salimath | Analog Devices |
| Ansell | Timothy | Google |
| Sundararaman | Ramesh | NVIDIA |
| Nunez-Rocha | Adrian | Qualcomm |
| Wieser | Jim | Texas Instruments |
| Hoofman | Romano | IMEC |
| Shalf | John | Lawrence Berkeley Labs |

# Appendix D: In-Person Workshop Agenda

| Friday May 20, 2022 | | |
|---|---|---|
| | | |
| **8:00AM** | | Continental Breakfast |
| **9:00AM** | Erik Brunvand, NSF | Welcome |
| **9:10AM** | Margaret Martinosi, NSF | Welcome |
| **9:20AM** | Matt Guthaus, UCSC | [Welcome and Overview](#) |
| **9:30AM** | Romano Hoofman, IMEC | [IMEC Overview & Q&A](#) |
| **10:15AM** | David Junkin, Cadence<br>Brandon Wang, Synopsys<br>Mohamed Kassem, Efabless<br>Tim Ansell, Google<br>Andrew Kahng, UCSD/OpenROAD | Panel: Proprietary vs Open-Source: Friends or Foes? |
| **11:15AM** | See Breakout Groups and Session Topics tabs | Breakout Session 1 |
| **12:00PM** | | Lunch |
| **12:45PM** | | Group Summaries |
| **1:30PM** | See Breakout Groups and Session Topics tabs | Breakout Session 2 |
| **2:15PM** | | Group Summaries |
| **3:00PM** | See Breakout Groups and Session Topics tabs | Breakout Session 3 |
| **3:45PM** | | Group Summaries |
| **4:30PM** | Matthew Guthaus | Organizing Working Groups and Chairs |
| | | Transit/Break |
| **6:30PM** | | Dinner @ Pacific Catch and Discussions |
| | | |
| Saturday May 21, 2022 | | |
| **8:00AM** | | Continental Breakfast |
| **9:00AM** | Matthew Guthaus | Opening Discussion |
| **9:30AM** | | Cross Cutting Discussions |
| **10:30AM** | | Working Group Breakouts |
| **12:00PM** | | Lunch & Closing Discussion |